\documentclass[11pt]{article}

\usepackage{amsmath}
\usepackage{amsmath}

\newcommand{\N}{N\raise.7ex\hbox{\underline{$\circ $}}$\;$}

\begin{document}

\thispagestyle{empty}
\title{Duffin-Kemmer-Petiau formalism  reexamined: non-relativistic
approximation   for spin 0 and spin 1 particles in  a Riemannian space-time}
\author{A.A. Bogush, V.V. Kisel,  N.G. Tokarevskaya, V.M. Red'kov
\\ Institute of Physics, National  Academy of Sciences of Belarus  }

\maketitle

\begin{abstract}

It is shown that the generally covariant Duffin-Kemmer-Petiau equation, formulated in the frame of
the Tetrode-Weyl-Fock-Ivanenko tetrad formalism,   allows for a non-relativistic approximation
if the  metric tensor  is of a special form.
The Pauli equation for a vector particle involves the Riemann  curvature tensor explicitly.
In analogous way, the procedure of the non-relativistic approximation in the theory
of scalar particle, charged and neutral, is investigated in the
background of Riemannian space-time. A generalized covariant
Schr\"{o}dinger equation is derived  when taking into account
non-minimal interaction term through scalar curvature $R(x)$,
 it substantially  differs from the conventional  generally  covariant Schr\"{o}dinger
equation produced when $R(x)=0$.
It is demonstrated that the the non-relativistic wave  function is always
complex-valued irrespective of the type of relativistic scalar
particle, charged or neutral, taken initially. The theory of
vector particle proves the same property: even if  the wave
function of the relativistic particle of spin 1 is  taken real,
the corresponding wave function in  the non-relativistic
approximation is complex-valued.

\end{abstract}

\section{Introduction}

 Matrix Duffin-Kemmer-Petiau formalism, for boson  fields
 has a long and rich  history inseparably linked with  description
of  photons and  mesons:

\begin{quotation}

\noindent
 Louis de  Broglie  [1-9],
 A. Mercier [10],
 G. Petiau [11],
 A. Proca  [12],
 Duffin [13],
N. Kemmer [14, 15],
H.J. Bhabha [16, 17],
F.J. Belinfante [18, 19],
S. Sakata --  M. Taketani
[20],
M.A. Tonnelat  [21],
 H.A.S. Erikson [22],
  E. Schr\"{o}dinger  [23-25],
  W. Heitler  [26, 27],
Yarish-Chandra [28-30],
 B. Hoffmann  [31],
  R. Utiyama [32],
  I.M. Gel'fand  --   A.M. Yaglom [33],
  J.A. Schouten [34],
  S.N. Gupta  [35],
  K. Bleuler [36],
  I. Fujiwara [37],
K.M. Case  [38],
H. Umezawa  [39],
 A.A.  Borgardt  [40-43],
 F.I. Fedorov [44],
 T. Kuohsien  [45],
 S. Hjalmars [46],
  A.A. Bogush -- F.I. Fedorov   [47],
   N.A. Chernikov  [48],
J. Beckers  [49],
P. Roman  [50],
F.I. Fedorov -- A.I. Bolsun [51],
 L. Oliver [52],
 J. Beckers,  C. Pirotte  [53],
 G. Casanova   [54],
 I.Yu. Krivski --  G.D. Romamenko -- V.I. Fushchych [55-57],
   A.A. Bogush et all . [58, 59],
    T. Goldman et al [60],
    R.A. Krajcik --  M. M. Nieto [61],
    J.D. Jenkins  [62, 63],
     E. Fischbach et al [64],
      K. Karpenko.  [65].

\end{quotation}

 Usually description of interaction between a quantum
mechanical particle and an~external classical  gravitational
field  looks  basically  differently for fermions and bosons (S. Weinberg [66]). For
a fermion, starting  from the  Dirac equation
\begin{eqnarray}
(\; i \gamma ^{a} \;  \partial _{a} \; - \;{  mc\over \hbar } \;) \; \Psi (x) \; =
\; 0
\nonumber
\end{eqnarray}

\noindent  we have to generalize through the~use of the~tetrad
formalism [66]. For a vector boson,  a~totally different approach is
generally  used:  it  consists  in  ordinary formal changing all
involved tensors and usual  derivative $\partial_{a}$ into
general relativity ones. For example, in  case  of a~vector
particle, the~flat space Proca equations
\begin{eqnarray}
 \partial _{a} \; \Psi _{b} \; - \;
\partial _{b} \; \Psi _{a} = \;{m c \over \hbar }  \Psi _{ab} \;\; , \qquad
\partial ^{b} \; \Psi _{ab} = {m c \over  \hbar} \;  \Psi  _{a}
\nonumber
\end{eqnarray}

\noindent being subjected to the~formal change $
\partial _{a} \; \rightarrow \;  \nabla _{\alpha } \; , \;
\Psi _{a} \; \rightarrow \; \Psi _{\alpha } ,\; \Psi _{ab} \;
\rightarrow  \; \Psi _{\alpha \beta } $ result in
\begin{eqnarray}
\nabla _{\alpha }\; \Psi _{\beta} - \nabla _{\beta }\;
\Psi_{\alpha } = {m c \over  \hbar} \; \Psi _{\alpha \beta } \; , \qquad \nabla
^{\beta }\; \Psi _{\alpha \beta } = {m c \over  \hbar} \; \Psi _{\alpha }
\label{1.1b}
\end{eqnarray}

\noindent
 However, the known Duffin-Kemmer-Petiau formalism in the curved space-time
till recent time was not  used, though such possibility is
known (S. Weinberg [66]).
The situation is changing now:

\begin{quotation}

\noindent
J.T. Lunardi --  B.M. Pimentel -- R.G. Teixeira --  J.S. Valverde  -- L.A. Manzoni   [67-69],
 V.Ya. Fainberg -- B.M. Pimentel [70],
 M. de Montigny --  F.C. Khanna --   A.E. Santana --  E.S.  Santos -- J.D.M. Vianna.
  [71], I.V. Kanatchikov[72],
  R. Casana -- V.Ya. Fainberg --   B.M.  Pimentel --  J.T.  Lunardi --  R.G.   Teixeira.
 [73, 74],
 Taylan Yetkin -- Ali Havare [75],
  S. Gonen, A. Havare, N. Unal  [76],
  A. Okninski  [77],
   Mustafa Salti, Ali Havare [78],
   A.A. Bogush --  V.S.  Otchik -- V.V. Kisel --   N.G. Tokarevskaya --   V.M. Red'kovl  [79-89],
.

\end{quotation}

\section{ Duffin-Kemmer-Petiau equation in gravitational field }

 We start from a flat
space equation  in  its  matrix  DKP-form
\begin{eqnarray}
(\; i\;  \beta ^{a} \; \partial_{a} \; - \; {m c \over  \hbar} \; )\;  \Phi  (x) =
0 \; ;
\label{2.1}
\end{eqnarray}

\noindent where
\begin{eqnarray}
\Phi  = ( \Phi _{0} , \; \Phi _{1} ,\; \Phi _{2}, \; \Phi _{3} ;
\; \Phi _{01}, \; \Phi _{02}, \; \Phi _{03}, \; \Phi _{23},\; \Phi
_{31}, \; \Phi _{12} )                     \;  ,
\nonumber
\\
\beta ^{a} = \left | \begin{array}{cc} 0 & \kappa ^{a} \\ \lambda
^{a} & 0
\end{array} \right | =
  \kappa  ^{a} \oplus \lambda  ^{a}  \;,
\;\; (\kappa  ^{a})_{j} ^{[mn]} \; = \; - i\; ( \delta ^{m}_{j} \;
g^{na} \;  - \; \delta
  ^{n}_{j} \; g^{ma} )\;\; ,
\nonumber
\\
( \lambda  ^{a})^{j}_{[mn]} \; = \;
  - i\; ( \delta  ^{a}_{m} \; \delta ^{j}_{n}  -
\delta ^{a}_{n} \;\delta ^{j}_{m} ) \; = - i\;\delta ^{aj}_{mn} \;
;
\label{2.2a}
\end{eqnarray}

\noindent $( g^{na} ) = \hbox{diag}( +1,-1,-1,-1 )$. The basic
properties of $\beta ^{a}$ are
\begin{eqnarray}
\beta ^{c} \; \beta ^{a} \; \beta ^{b} = \left| \begin{array}{cc}
     0 & \kappa ^{c} \;\lambda ^{a}\;\kappa ^{b} \\
\lambda^{c} \; \kappa ^{a} \; \lambda ^{b} & 0
\end{array} \right | ,  \qquad
(\lambda ^{c} \; \kappa ^{a} \; \lambda ^{b}) ^{j}_{[mn]} =
 i\; ( \; \delta ^{cb}_{mn} \; g^{aj} \;  - \; \delta ^{cj}_{mn} \;  g^{ab}\; ) \; ,
\nonumber
\\
(\kappa ^{c} \; \lambda ^{a} \; \kappa ^{b})^{[mn]}_{j}  =
 i \; [\; \delta ^{a}_{j}\; (g^{cm} \; g^{bn} \; - \; g^{cn}\; g^{bm} ) \; + \;
\;g^{ac}\; ( \delta ^{n}_{j} \; g^{mb} \;  - \; \delta ^{m}_{j}\;
g^{nb} ) \;] \; ,
\label{2.2b}
\end{eqnarray}

\noindent and
\begin{eqnarray}
 \beta ^{c} \; \beta ^{a} \; \beta  ^{b} \;  + \;
\beta  ^{b} \; \beta ^{a} \; \beta ^{c}  =  \beta ^{c} \; g^{ab}
\; + \; \beta ^{b} g^{ac}\;  ,
\nonumber
\\
\; [\beta ^{c} , j^{ab} ] =  g^{ca} \;\beta^{b} \; - \; g^{cb} \;
\beta ^{a}\;   , \qquad j^{ab} =  \beta ^{a} \; \beta ^{b} \; - \;
\beta ^{b} \; \beta ^{a}\;  ,
\nonumber
\\
\;[j^{mn}, j^{ab}]  = ( \;g^{na} \; j^{mb} \;  - \; g^{nb} \; j^{ma}
\;) \; - \; (\; g^{ma} \;j^{nb}\;   - \; g^{mb} \; j^{na}\; )\;  .
\label{2.2c}
\end{eqnarray}

In accordance with tetrad recipe one  should generalize the
DKP-equation as follows
\begin{eqnarray}
[ \; i \; \beta ^{\alpha }(x)\; (\partial_{\alpha} \;  +  \;
B_{\alpha }(x) ) \; - {m c \over  \hbar} \;  ] \;\Phi  (x)  = 0 \; ,
\nonumber
\\
\beta ^{\alpha }(x) = \beta ^{a} e ^{\alpha }_{(a)}(x) \; , \;\;
B_{\alpha }(x) = {1 \over 2}\; j^{ab} e ^{\beta }_{(a)}\nabla
_{\alpha }( e_{(b)\beta })  \;  .
\label{2.3}
\end{eqnarray}

\noindent This equation  contains  the  tetrad $e^{\alpha
}_{(a)}(x)$ explicitly. Therefore, there must  exist
a~possibility  to  demonstrate
 the~equivalence of  any  variants  of  this  equation
associated with various tetrads:
\begin{eqnarray}
e^{\alpha }_{(a)}(x)  \;\; , \qquad   e'^{\alpha }_{(b)}(x)\; = \;
 L^{\;\;b}_{a} (x) \; e^{\alpha }_{(b)}(x)  \; ,
\label{2.4a}
\end{eqnarray}

\noindent   $L^{\;\;b}_{a} (x)$  is a local Lorentz  transformation. We will  show  that  two such equations
\begin{eqnarray}
 [\;  i \beta
^{\alpha }(x) \; (\partial_{\alpha}  + B_{\alpha }(x)) \;  - \; {m c \over  \hbar}
\;] \; \Phi  (x)  = 0 \; ,
\nonumber
\\
\; [\; i \beta'^{\alpha }(x) \; (\partial_{\alpha}  + B'_{\alpha
}(x))  - {m c \over  \hbar} \; ]\;  \Phi'(x)  = 0 \; ,
\label{2.4b}
\end{eqnarray}

\noindent generating in tetrads $e^{\alpha }_{(a)}(x)$    and
 $e'^{\alpha }_{(b)}(x)$
respectively,  can  be  converted  into  each  other  through  a local gauge
transformation:
\begin{eqnarray}
\Phi '(x) = \left | \begin{array}{c}
                   \phi'_{a}(x) \\ \phi'_{[ab]}(x)
\end{array} \right | =
\left | \begin{array}{cc}
         L_{a}^{\;\;l} & 0 \\
         0 &  L_{a}^{\;\;m} L_{b}^{\;\;n}
\end{array} \right | \;\;
\left | \begin{array}{c}
                   \phi_{l}(x) \\ \phi_{[mn]}(x)
\end{array} \right |      \; .
\label{2.4c}
\end{eqnarray}

\noindent Starting from the first equation in (\ref{2.4b}), let us
obtain an equation for $\Phi'(x)$. Allowing for $\Phi(x) = S(x)\;
\Phi (x)$, we get
\begin{eqnarray}
[\; i\; S \; \beta ^{\alpha } \; S^{-1} (\partial _{\alpha} \; +
\; S \; B_{\alpha }\; S^{-1} \; + \; S \; \partial_{\alpha}\;
S^{-1}) \; -
 \; {m c \over  \hbar} \;] \; \Phi'(x) = 0 \; .
\nonumber
\end{eqnarray}

\noindent One should  verify   relationships
\begin{eqnarray}
S(x) \; \beta ^{\alpha }(x) \; S^{-1}(x) = \beta'^{\alpha}(x) \; ,
\;
\label{2.5a}
\\
 S(x) \; B_{\alpha}(x)\; S^{-1}(x) \; + \; S(x) \; \partial_{\alpha}\; S^{-1}(x)  =
 B'_{\alpha} (x)      \; .
\label{2.5b}
\end{eqnarray}

\noindent The first one can be rewritten as
\begin{eqnarray}
S(x) \; \beta ^{a} \; e ^{\alpha }_{(a)}(x) \; S^{-1}(x) \; =
\beta ^{b} \; e'^{\alpha }_{(b)}(x)   \; ;
\nonumber
\end{eqnarray}

\noindent from where we come to
\begin{eqnarray}
S(x) \;\beta ^{a} \; S^{-1}(x) \; = \; \beta ^{b} \;
L^{\;\;a}_{b}(x) \; .
\label{2.5c}
\end{eqnarray}

\noindent The latter condition is well-known in $DKP$-theory;  one can verify it through the use of the block   structure
of $\beta ^{a}$, which provides two  relations:
\begin{eqnarray}
L(x) \; \kappa ^{a} \; [\; L^{-1}(x) \otimes  L(x)^{-1}\;] \; =
 \;\kappa ^{b} \; L^{\;\;a}_{b}(x)  \;,
 \qquad [\; L(x) \otimes  L(x)\; ] \; \lambda ^{a} \; L(x)^{-1} \; = \;
\lambda ^{b} \; L^{\;\;a}_{b}(x) \; .
\nonumber
\end{eqnarray}

\noindent They  will be satisfied identically, after we take
explicit form of $\kappa ^{a}$ and $\lambda ^{a}$  and   allow for
the  pseudo orthogonality condition: $ g^{al} \;
(L^{-1})^{\;\;k}_{l}(x) = g^{kb}\; L^{\;\;a}_{b}(x)\; . $ Now, let
us pass to the proof of the relationship (\ref{2.5b}).  By using the
determining relation for $DKP$- connection we readily produce
\begin{eqnarray}
S(x)\; \partial_{\alpha} \; S^{-1}(x) \; = \;  \; B'_{\alpha }(x)
\; - \; {1 \over 2} \; j^{mn} L^{\;\;n}_{m}(x)\; g_{ab} \;
\partial _{\alpha}\;
 L^{\;\;b}_{n}(x) \; \;
\nonumber
\end{eqnarray}

\noindent therefore  eq. (\ref{2.5b})   results in
\begin{eqnarray}
S(x) \; \partial_{\alpha} \; S^{-1}(x) =  \; {1\over2}\;
L^{\;\;a}_{m}(x)\; g_{ab}\; (\; \partial_{\alpha }
\;L^{\;\;b}_{n}(x)\; ) \; .
\nonumber
\end{eqnarray}

\noindent The latter condition is an identity  readily verified
through the~use of block  structure of all involved matrices.
Thus, the equations from (\ref{2.4b}) are  translated  into  each other.
In other words, they  manifest  a~gauge  symmetry under local
Lorentz transformations in  complete analogy with more familiar
Dirac particle case. In  the~same time, the~wave function  from
this  equation  represents  scalar quantity relative to general
coordinate transformations: that is, if $\; x^{\alpha } \;
\rightarrow  \; x'^{\alpha } = f^{\alpha }(x)$ ,  then $\;
\Phi'(x) = \Phi (x)$.

It remains to demonstrate that this $DKP$  formulation  can  be
inverted into the Proca formalism in terms of  general  relativity
tensors. To this end, as a~first  step,  let  us  allow  for
the~sectional structure of $\beta ^{a}, J^{ab}$  and $\Phi (x)$ in
the~$DKP$-equation; then instead of (\ref{2.3}) we get
\begin{eqnarray}
i \; [\; \lambda^{c} \; e^{\alpha }_{(c)} \; (\; \partial_{\alpha}
\; + \; \kappa ^{a} \; \lambda ^{b} \; e^{\beta }_{(a)} \;
\nabla_{\alpha }\; e_{(b)\beta }\; ) \;
]^{\;\;\;\;\;\;\;l}_{[mn]}\; \Phi _{l} = {m c \over  \hbar}\; \Phi _{[mn]}    \; ,
\nonumber
\\
i \; [\kappa ^{c} \; e^{\alpha }_{(c)} \; ( \;\partial_{\alpha} \;
+ \; \lambda ^{a} \; \kappa ^{b}\; e^{\beta }_{(a)} \; \nabla
_{\alpha } \; e_{(b)\beta }\; ) \; ]^{\;\;\;[mn]}_{l} \;
\Phi_{[mn]}  = {m c \over  \hbar} \Phi _{l} \; ,
\label{2.6a}
\end{eqnarray}

\noindent which lead to
\begin{eqnarray}
 (e_{(a)}^{\alpha} \; \partial_{\alpha} \; \Phi _{b} \; - \;
e^{\alpha }_{(b)} \; \partial_{\alpha}\; \Phi _{a}) \; + \; (
\gamma ^{c}_{\;\;ab} - \gamma ^{c}_{\;\;ba} ) \; \Phi _{c} = {m c \over  \hbar} \;
\Phi _{ab}\; ,
\nonumber
\\
e^{(b)\alpha } \;\partial_{\alpha } \; \Phi _{ab} \; + \; \gamma
^{nb} _{\;\;\;\;n} \Phi _{ab} + \gamma ^{\;\;mn}_{a} \Phi _{mn}  =
{m c \over  \hbar} \; \Phi _{a}  \; ;
\label{2.6b}
\end{eqnarray}

\noindent the  symbol $\gamma _{abc}(x)$ is used to designate
Ricci  coefficients: $ \gamma _{abc}(x) \; =  \; - \; e_{(a)\alpha
; \beta  } \; e^{\alpha }_{(b)} \; e^{\beta }_{(c)} \; . $ In
turn,  (\ref{2.6b}) will look  as   the~Proca equations (\ref{1.1b}) after
they are rewritten in terms of tetrad  components
\begin{eqnarray}
\Phi _{a}(x) \;= \; e^{\alpha }_{(a)} (x) \; \Phi _{\alpha }(x) ,
\qquad \Phi _{ab}(x)= e^{\alpha }_{(a)} (x) \;  e^{\beta
}_{(b)}(x) \; \Phi_{\alpha \beta }(x)     \; .
\label{2.7}
\end{eqnarray}

So, as evidenced by the~above, the~manner of introducing
the~interaction between a~spin  $1$  particle  and  external
classical gravitational field can be successfully unified
with~the approach  that occurred with regard to a~spin $1/2$
particle and was first  developed by  Tetrode,  Weyl,  Fock,
Ivanenko.  One  should   attach   great significance to that
possibility of unification.  Moreover,  its absence would  be
a~very strange fact.  Let  us  add some  more details.

The manner of extending the~flat  space  Dirac  equation  to
general relativity case indicates clearly that the~Lorentz  group
underlies equally both these theories. In other words, the~Lorentz
group retains its importance and significance  at  changing  the
Minkowski space  model  to  an~arbitrary  curved  space-time.  In
contrast to this,  at  generalizing  the~Proca   formulation,  we
automatically destroy any relations to the~Lorentz group, although
the~definition itself for a~spin $1$ particle as an~elementary
object was  based on    this  group.  Such  a~gravity
sensitiveness to the~fermion-boson division might appear rather
strange  and  unattractive asymmetry, being subjected to
the~criticism. Moreover,  just  this feature  has brought about
a~plenty  of  speculation  about  this matter. In any case, this
peculiarity of particle-gravity  field interaction  is  recorded
almost  in  every  handbook.

\section{ Non-relativistic approximation,  10-component formalism }

The first who was interested in non-relativistic equation for a
particle with spin 1 was  A. Proca [12]. Let us  consider such a
problem in presence of  external gravitational fields. To have a non-relativistic approximation,
we must  use
the limitation on space-time models:
\begin{eqnarray}
dS^{2} = (dx^{0})^{2} + g_{ij}(x) dx^{i} dx^{j}  \; , \qquad
e_{(a)\alpha}(x) = \left | \begin{array}{cc} 1  &  0  \\ 0  &
e_{(i)k}(x)  \end{array} \right |  .
 \label{3.1}
\end{eqnarray}

\noindent DKP-equation in presence both of curved space background
 and electromagnetic field is
\begin{eqnarray}
[\; i \beta^{0} \; D_{0} + i    \beta^{k}(x)\;  D_{k}   -  {mc
\over \hbar}\; ] \;  \Psi  = 0\;  ,
\nonumber
\\
D_{\alpha} =  \partial _{\alpha} + B_{\alpha} (x) - i {e\over \hbar
c} A_{\alpha}(x) \; .
\label{3.2}
\end{eqnarray}

\noindent In the  metric  (\ref{3.1}), expressions for vector
connections become much simpler, indeed
\begin{eqnarray}
B_{0} = {1 \over 2 } J^{ik} e_{(i)}^{m}  (\nabla_{0}
e_{(k) m }) \; , \qquad  B_{l} = {1 \over 2 } J^{ik} e_{(i)}^{m}
(\nabla_{l} e_{(k) m })  \;,
\label{3.3}
\end{eqnarray}

\noindent so there is no contribution from $J^{0k}$ generators.
Because of identities
\begin{eqnarray}
  \beta^{0} \beta^{0} J^{kl} = J^{kl}
\beta^{0} \beta^{0} ] \qquad \Longrightarrow \qquad \beta^{0}
\beta^{0} B_{\alpha} (x) = B_{\alpha}(x) \beta^{0} \beta^{0}
\label{3.5}
\end{eqnarray}

\noindent  the operator   $D_{k}$ commutes with
$(\beta^{0})^{2}$.  Multiplying eq.  (\ref{3.2}) by projective
$(\beta^{0})^{2}$ and $1-(\beta^{0})^{2}$ , and taking into
account  relations
\begin{eqnarray}
(\beta^{0})^{2} \beta^{l}(x) = \beta^{l}(x) \; [ 1 -
(\beta^{0})^{2} ]  \;,
\nonumber
\\
\;[ \;1 - (\beta^{0})^{2} \; ] \;   \beta^{l}(x) = \beta^{l}(x)
(\beta^{0})^{2} \;  ,\qquad (\beta^{0})^{3}= \beta^{0} \; ,
\label{3.6}
\end{eqnarray}

\noindent we get to equations for  $\chi$ and $\varphi$:
\begin{eqnarray}
 \chi = ( \beta^{0}) ^{2} \Psi \; , \qquad
\varphi = (1  - (\beta^{0})^{2}) \Psi \;  , \qquad \Psi = \chi +
\varphi \;,
\nonumber
\\
i \beta^{0} \; D_{0} \chi   +   i \beta^{k}(x) \; D_{k}  \varphi =
  {mc \over \hbar}\;  \chi  \; ,
\qquad i \beta^{k}(x) \; D_{k} \chi  = { mc \over \hbar } \;  \varphi \;
.
\label{3.7}\
\end{eqnarray}

\noindent Excluding  a non-dynamical part  $\varphi$, we arrive at
\begin{eqnarray}
i \beta^{0}\; D_{0} \chi - {\hbar \over mc}\;  \beta^{k}(x)
\;\beta^{l}(x) \; D_{k}  D_{l} \;  \chi = { mc \over \hbar}\; \chi
\; .
\label{3.8}
\end{eqnarray}

 Now let us introduce  two operators
\begin{eqnarray}
\Pi_{\pm} = {1 \over 2 }\;  \beta^{0} \;( 1 \pm \beta^{0})\; , \qquad
 \Pi_{+} \beta^{0} = + \; \Pi_{+}  \;, \qquad
\Pi_{-} \beta^{0} = - \; \Pi_{-} \; .
\nonumber
\end{eqnarray}

\noindent From  (\ref{3.8})  it follows
\begin{eqnarray}
i D_{0} \;  \Pi _{+} \;\chi   - {\hbar \over m c} \; \Pi_{+}  \;
\beta^{k}(x) \beta^{l}(x)  \; D_{k} D_{l}  \; \chi  - {mc \over
\hbar} \; \Pi_{+} \;\chi  = 0  \; ;
\label{3.10}
\end{eqnarray}

\noindent with the help of
\begin{eqnarray}
\Pi_{+} \; \beta^{k}(x) \beta^{l} (x) = {1 \over 2}  \; [\; (-
\beta^{l}(x) \beta^{k}(x)  +  g^{lk}(x)  ) \;   \beta^{0}  +
 \beta^{k}(x) \beta^{l}(x) \;   (\beta^{0}) ^{2}  \; ] \; ,
\nonumber
\end{eqnarray}

\noindent eq.  (\ref{3.10}) leads to
\begin{eqnarray}
i   D_{0}   \; ( \Pi _{+} \chi )  - {\hbar \over 2m c} \;
 [\;
(- \beta^{l} (x) \;\beta^{k} (x)  +  g^{lk}(x) ) \;  \beta^{0}  +
\nonumber
\\
+
 \beta^{k}(x)\; \beta^{l}(x)  \;  (\beta^{0}) ^{2} \;  ]\;
D_{k} D_{l}  \;\chi  = {mc \over \hbar}   \;(\Pi_{+} \chi ) \; .
\label{3.11}
\end{eqnarray}

\noindent In the same manner, starting from
\begin{eqnarray}
- i   D_{0} \; \Pi_{-} \;\chi - { \hbar \over mc} \; \Pi_{-}  \;
\beta^{k} (x) \;\beta^{l} (x)  \; D_{k} D_{l}  \; \chi = {mc \over
\hbar}  \; \Pi_{-} \; \chi \;  ,
\nonumber
\end{eqnarray}

\noindent with the help of
\begin{eqnarray}
\Pi_{-} \; \beta^{k}(x) \beta^{l}(x) = {1 \over 2} \;   [\;   (-
\beta^{l}(x) \beta^{k}(x)  + g^{kl}(x)) \; \beta^{0} -
\beta^{k}(x) \beta^{l}(x) \;  (\beta^{0})^{2}  \; ]\;  ,
\nonumber
\end{eqnarray}

\noindent we get
\begin{eqnarray}
- i   D_{0} \; \Pi_{-} \chi - { \hbar \over 2mc}\; [\;  (-
\beta^{l}(x)\; \beta^{k}(x) + g^{kl}(x) ) \;  \beta^{0} -
\nonumber
\\
- \beta^{k} (x) \;\beta^{l}(x)\;  (\beta^{0})^{2}  \;  ]\; D_{k}
D_{l} \;  \chi  =  {mc \over \hbar} \;  \Pi_{-} \; \chi  \; .
\label{3.14}
\end{eqnarray}

\noindent  Changing matrices  $\beta^{0} $ and $(\beta^{0})^{2}$
by
\begin{eqnarray}
\Pi_{+} + \Pi_{-} = \beta^{0}  , \qquad \Pi_{+} - \Pi_{-} =
\beta^{0} \beta^{0}\; ,
\nonumber
\end{eqnarray}

\noindent and  using the notation ($\Pi_{-} \chi = \chi_{-}  , \; \Pi_{+} \chi = \chi_{+} $)
\begin{eqnarray}
J^{[kl] }(x) = \beta^{k} (x)\;\beta^{l}(x) -  \beta^{l}(x)
\;\beta^{k}(x) \; , \qquad
J^{(kl) }(x) = \beta^{k}(x) \;\beta^{l}(x) +  \beta^{l} (x)\;
\beta^{k}(x)  \; ,
\nonumber
\end{eqnarray}

\noindent  reduce eqs.  (\ref{3.11}) and (\ref{3.14}) to the form
\begin{eqnarray}
i D_{0}  \;  \chi _{+}   -   { \hbar \over 2mc} \; [\; J^{[kl]}(x)
\;  D_{k} D_{l}  \; \chi_{+}   - J^{(kl)}(x)  \; D_{k} D_{l}  \;
\chi _{-}   +
 D^{l} D_{l} \;  (\chi_{+} + \chi_{-}) \;] =   {mc \over \hbar }
\;\chi_{+}  \; ,
\nonumber
\\
- i  D_{0} \;   \chi _{-}   -  { \hbar \over 2mc}\;  [\; J^{[kl]}
(x)\;  D_{k} D_{l} \; \chi_{-}  -
 J^{(kl)} (x) \; D_{k} D_{l}  \;\chi _{+}  +
 D^{l} D_{l}   \; (\chi_{+} + \chi_{-})\;  ] =  {mc \over \hbar }
\; \chi_{-} \;  .
\nonumber
\\
\label{3.15}
\end{eqnarray}

 Now, one should  separate a special factor depending on
the rest-energy, so that in eq.  (\ref{3.15}) one should make one formal change:
\begin{eqnarray}
\chi  \;\; \Longrightarrow \;\; \exp(-i{mc^{2} \over \hbar} t)\;
\chi \;, \qquad
i D_{0} \qquad \Longrightarrow  \qquad i  D_{0}   +   {mc \over
\hbar }\;   . \label{3.16}
\end{eqnarray}

\noindent As a result, eq.  (\ref{3.15}) gives
\begin{eqnarray}
i D_{0} \;  \chi _{+}  -  { \hbar \over 2mc} \; [\; ( J^{[kl]}(x)
\;  D_{k} D_{l} \;  \chi_{+}   -
 J^{(kl)}(x) \; D_{k} D_{l} \; \chi _{-} ) + D^{l} D_{l} \;
(\chi_{+} + \chi_{-}) \; ] = 0  \; ,
\nonumber
\\
- i D_{0} \;  \chi _{-}  -  { \hbar \over 2mc}  \; [\; (
J^{[kl]}(x) \; D_{k} D_{l}\;  \chi_{-}   -
 J^{(kl)}(x)\;  D_{k} D_{l} \; \chi _{+} )  + D^{l} D_{l}\;
(\chi_{+} + \chi_{-}) \; ] = 2 { m c \over \hbar } \; \chi_{-}  \;
.
\nonumber
\\
\label{3.17}
\end{eqnarray}

Now, taking $\chi_{-}$ as small and ignoring the term  $- i D_{0}
\chi _{-}$ compared with
 ${mc \over \hbar } \chi_{-}$ we arrive at
\begin{eqnarray}
(  J^{(kl)}  \; D_{k} D_{l}     - D^{l} D_{l} )  \;  \chi_{+}  =
{4 m^{2} c^{2} \over \hbar^{2} }\;  \chi_{-}  \; ,
\nonumber
\\
i \hbar D_{t}  \; \chi _{+} =   { \hbar^{2}  \over 2m} \; (  D^{l}
D_{l}  + J^{[kl]} \; D_{k} D_{l}    )\;   \chi_{+} \;  ,
\label{3.18}
\end{eqnarray}

\noindent The second  equation in (\ref{3.18})  should be considered as a
non-relativistic Pauli equation for spin 1 particle in
DKP-approach.

It is interesting to see what is the form of the non-relativistic  approximation   in
tensor form?
 At first, let us restrict ourselves to the case of the
flat space. From   (\ref{3.7}) it follows
\begin{eqnarray}
\beta^{0}  \left | \begin{array}{l}
\Phi_{0}  \\  \Phi_{1}  \\  \Phi_{2}  \\  \Phi_{3}  \\
\Phi_{01}  \\ \Phi_{02}  \\ \Phi_{03}  \\ \Phi_{23}  \\ \Phi_{31}
\\ \Phi _{12}
\end{array} \right |  =
\left | \begin{array}{c}
0  \\ i \Phi_{01}  \\ i\Phi_{02}  \\ i\Phi_{03}  \\
-i \Phi_{1}  \\  -i \Phi_{2}  \\ - i\Phi_{3}  \\ 0  \\ 0  \\ 0
\end{array} \right |\; , \qquad
\chi = ( \beta^{0}) ^{2} \;\Psi  = \left | \begin{array}{c}
0  \\ \Phi_{1} \\ \Phi_{2}  \\ \Phi_{3}  \\
\Phi_{01}  \\ \Phi_{02}  \\ \Phi_{03} \\ 0   \\ 0 \\ 0
\end{array} \right | \; , \;\;\; \varphi = (1  - (\beta^{0})^{2})
\; \Psi = \left | \begin{array}{c} \Phi_{0} \\  0  \\  0  \\  0
\\ 0  \\ 0  \\  0  \\ \Phi_{23}  \\ \Phi_{31}  \\ \Phi_{12}
\end{array} \right |
\; ,
\nonumber
\end{eqnarray}

\noindent so that
\begin{eqnarray}
\chi_{+} =  {1 \over 2} \; \left | \begin{array}{c}
0  \\
(\Phi_{1}\; + \;i \;\Phi_{01}) \\
(\Phi_{2}\; +\; i \; \Phi_{02}) \\
(\Phi_{3} \; +\; i \; \Phi_{03}) \\
-i (\Phi_{1} + i \Phi_{01}) \\
-i (\Phi_{2} + i \Phi_{02}) \\
-i (\Phi_{3} + i \Phi_{03}) \\
0 \\  0  \\  0   \end{array} \right | \; , \;\;\; \chi_{-} = {1
\over 2} \; \left | \begin{array}{c}
0  \\
-(\Phi_{1}\; -\;i \;\Phi_{01}) \\
-(\Phi_{2}\; -\; i \; \Phi_{02}) \\
-(\Phi_{3} \; -\; i \; \Phi_{03}) \\
-i (\Phi_{1} - i \Phi_{01}) \\
-i (\Phi_{2} - i \Phi_{02}) \\
-i (\Phi_{3} - i \Phi_{03}) \\
0 \\  0  \\  0   \end{array} \right | .
\nonumber
\end{eqnarray}

\noindent Instead of $\Phi_{k}$ и $\Phi_{0k}$, let us  introduce
new  field variables:
\begin{eqnarray}
{1 \over 2} \;(  \Phi_{k} \;-\; i\; \Phi_{0k} ) = M_{k} \; ,
\qquad {1 \over 2} \;(  \Phi_{k} \; +\; i\; \Phi_{0k} ) = B_{k} \; ,
\nonumber
\\
\Phi_{k} = B_{k} \; + \; M_{k} \; , \qquad \Phi_{0k} = -i \; (
B_{k}\; - \; M_{k} ) \; ;
\label{3.20}
\end{eqnarray}

\noindent that is
\begin{eqnarray}
\chi_{+} =   \left | \begin{array}{c}
0  \\
\vec{B} \\
-i\; \vec{B}  \\
0 \\  0  \\  0   \end{array} \right | \; , \qquad \chi_{-} =
 \left | \begin{array}{c}
0        \\
- \vec{M}  \\
-i \; \vec{M} \\
0 \\  0  \\  0   \end{array} \right |\; .
\label{3.21}
\end{eqnarray}

\noindent Thus, in tensor representation the big and small
components coincides with 3-vectors  $\vec{B}$ and $\vec{M}$
respectively. Now it is a matter of simple calculation to repeat
the limiting procedure in tensor basis. Indeed, starting from
Proca equations (it is convenient to change the notation $mc/ \hbar \Longrightarrow m$)
\begin{eqnarray}
D_{0}\; \Phi_{k} \;-\; D_{k}\;  \Phi_{0} = m \;\Phi_{0k} \; ,
\;\;\; D_{k} \;\Phi_{l} \; - \; D_{l} \; \Phi_{k} = m\;  \Phi_{kl}
\; ,
\nonumber
\\
D^{l} \; \Phi_{0l} = m \;\Phi_{0} \; , \qquad D^{0} \; \Phi_{k0}\;
+ \; D^{l} \; \Phi_{kl} = m\; \Phi_{k} \; ,
\label{3.22}
\end{eqnarray}

\noindent and excluding the non-dynamical components  $\Phi_{0},
\; \Phi_{kl}$,
\begin{eqnarray}
D_{0}\; \Phi_{k} \;- \;{1 \over m} \; D_{k} \; D^{l} \; \Phi_{0l}
= m \; \Phi_{0k} \; ,
\nonumber
\\
D^{0} \; \Phi_{k0} \; +\;  {1 \over m } \; D^{l} \; ( D_{k}
\;\Phi_{l} \; -\; D_{l} \;\Phi_{k}) = m \; \Phi_{k} \; .
\label{3.23}
\end{eqnarray}

\noindent and further
\begin{eqnarray}
m\; (\Phi_{k} \; \pm\;  i \; \Phi_{0k} ) = ( \; D^{0} \; \Phi_{k0}
\; +\; {1 \over m } \; D^{l}  \; D_{k} \; \Phi_{l} \; -
\nonumber
\\
- \; {1 \over m } \; D^{l} \;D_{l} \; \Phi_{k}\; ) \; \pm i \; (\;
D_{0} \; \Phi_{k}\; -\; {1 \over m} \; D_{k} \; D^{l} \; \Phi_{0l}
\; ) \; .
\label{3.24}
\end{eqnarray}

\noindent From these, with the help of
 (\ref{3.20}), we get to
\begin{eqnarray}
2 m\; B_{k} = +2 i \;D_{0} B_{k} - {1 \over m } \; D^{l}D_{l}
(B_{k} \;+ \;M_{k} ) \; +
\nonumber
\\
+ \; {1 \over m } \; [ \; (D^{l}\; D_{k} \; - \; D_{k} \; D^{l} )
\; B_{l} \; + \; (D^{l} \; D_{k} \; +  \; D_{k} \; D^{l} ) \;
M_{l} \; ] \; ,
\nonumber
\\
2 m \; M_{k} =  - 2 i \; D_{0} \; M_{k} \; - {1 \over m } \; D^{l}
\; D_{l} \; (B_{k}\; + \; M_{k} ) \; +
\nonumber
\\
+ \; {1 \over m } \; [ \; ( D^{l} \; D_{k} \; + \; D_{k} \; D^{l}
) \; B_{l} \; + (D^{l} \; D_{k}\; -\; D_{k}\; D^{l} ) \; M_{l} \;
] \; .
\label{3.26}
\end{eqnarray}

\noindent After separating the rest-energy term
\begin{eqnarray}
i\; D_{0} \; B_{k} \; \Longrightarrow \;  (i\; D_{0} \; + \; m)\;
B_{k} \; , \qquad i \; D_{0} \; M_{k} \; \Longrightarrow  \; (i\;
D_{0} \; + \; m) \;M_{k}\; ;
\nonumber
\end{eqnarray}

\noindent  from (\ref{3.26}) we arrive at
\begin{eqnarray}
+ i \;D_{0}\; B_{k} \; -\; {1 \over  2 m } \; \{ \;  D^{l}\; D_{l}
\; (B_{k} \; +\; M_{k} ) \; +
\nonumber
\\
+ \;   (D^{l} \; D_{k} \; -\; D_{k}\; D^{l} ) \; B_{l} \; + (D^{l}
\; D_{k} \; + \; D_{k} \; D^{l} )\; M_{l} \; \}  = 0 \; ,
\nonumber
\\
-  i \; D_{0} \; M_{k} \; - {1 \over 2 m } \;\{ \;  D^{l}\; D_{l}
\; (B_{k} \; +\; M_{k} ) \; +
\nonumber
\\
+ ( D^{l} \; D_{k} \; +\;  D_{k} \; D^{l} )\; B_{l} \; + (D^{l} \;
D_{k} \; - \; D_{k} \; D^{l} ) \; M_{l} \; \}  =  4m \; M_{k} \; .
\label{3.27}
\end{eqnarray}

 Therefore, a non-relativistic wave equation for the big
component  $\vec{B}$  has the form (let us change the  notation:
$B_{k}(x) \Longrightarrow \psi_{k}(x)$)
\begin{eqnarray}
+ i \;D_{0} \; \psi_{k} = {1 \over  2 m } \; [ \;-  D_{l}\; D_{l}
\; \psi_{k}  \; -
 \;   (D_{k} \; D_{l} \; - \; D_{l} \; D_{k} ) \; \psi_{l} \;  ]  \; .
\label{3.28}
\end{eqnarray}

\section{ Tetrad  3-dimensional non-relativistic equation}

The non-relativistic equation  (\ref{3.18}) in DKP-formalism is
symbolical in a sense, because it is written formally for a 10-component
function  though in fact the non-relativistic function is a  3-vector. Let us turn to the above
limiting procedure again (for shortness  $e/\hbar c
\Longrightarrow e,\; mc/ \hbar \Longrightarrow m $)
\begin{eqnarray}
[\; i \beta^{0}\; D_{0} + i    \beta^{l}(x) \;D_{l}   -  m \;  ]\;
\Psi  = 0 \; ,
\label{4.1}
\end{eqnarray}

\noindent where
\begin{eqnarray}
\beta^{l}(x) = \beta^{k}  e_{(k)}^{l}(x) \;  , \qquad D_{l} =
\partial_{l} + B_{l} - i e A_{l} \;, \qquad D_{0} = \partial_{0} +
B_{0} - i e A_{0} \; ,
\nonumber
\\
B_{0} = {1 \over 2 } J^{ik} e_{(i)}^{m}\; (\nabla_{0}  e_{(k) m })
\;, \qquad B_{l} = {1 \over 2 } J^{ik} e_{(i)}^{m}\;  (\nabla_{l}
e_{(k) m })\;  . \label{4.2}
\end{eqnarray}

\noindent With the  use of block-form for DKP-matrices
\begin{eqnarray}
\beta^{0} = \left | \begin{array}{cccc}
0  &  0  &  0  &  0  \\
0  &  0  &  i_{3}  &  0  \\
0  &  -i_{3} &  0  &  0  \\
0  &  0  &  0  &  0 \end{array} \right | \; , \qquad \beta^{k} =
\left | \begin{array}{cccc}
0  &  0  &  w_{k} &  0  \\
0  &  0  &  0  &  \tau_{k}  \\
\tilde{w}_{k}  &  0 &  0  &  0  \\
0  &  -\tau_{k}  &  0  &  0 \end{array} \right | \; , \label{4.3}
\end{eqnarray}

\noindent where
\begin{eqnarray}
_{1} = (i , 0 , 0 ) \;, \qquad \qquad  w_{2} = (0 , i , 0 ) \;, \qquad  \qquad
w_{3} = (0,  0 , i )\;  ,
\nonumber
\\
\tilde{w}_{1} = \left | \begin{array}{c}  i \\  0 \\  0
\end{array} \right|  , \qquad \tilde{w}_{2} = \left |
\begin{array}{c} 0 \\  i \\  0 \end{array} \right |  , \qquad
\tilde{w}_{3} = \left | \begin{array}{c} 0 \\  0 \\  i \end{array}
\right |  , \qquad i_{3} = \left | \begin{array}{ccc} i  &  0  & 0
\\  0  &  i  &  0  \\ 0  &  0  &  i  \end{array} \right | ,
\nonumber
\\
\tau_{1} = \left | \begin{array}{ccc}
0  &  0  &  0  \\
0  &  0  & -i  \\
0  &  i  &  0    \end{array} \right |  ,\qquad \tau_{2} = \left |
\begin{array}{ccc}
0  &  0  &  i  \\
0  &  0  &  0  \\
-i  & 0  &  0    \end{array} \right |  , \qquad \tau_{3} = \left |
\begin{array}{ccc}
0  &  -i  &  0  \\
i  &  0  & 0  \\
0   &  0  &  0    \end{array} \right |   \label{4.4}
\end{eqnarray}

\noindent and taking explicit  form of generators
 $J^{kl}$, for connections $B_{0}(x)$  and   $B_{l}(x)$  we have expressions
\begin{eqnarray}
B_{0}(x) = \left | \begin{array}{cccc}
0  &  0        &    0      &  0  \\
0  &  b_{0}(x) &    0      &  0  \\
0  &  0        & b_{0}(x)  &  0  \\
0  &  0        &    0      & b_{0}(x)      \end{array} \right |
,\qquad B_{l}(x) = \left | \begin{array}{cccc}
0  &  0        &    0      &  0  \\
0  &  b_{l}(x) &    0      &  0  \\
0  &  0        & b_{l}(x)  &  0  \\
0  &  0        &    0      & b_{l}(x)      \end{array} \right | ,
\label{4.5}
\end{eqnarray}

\noindent where
\begin{eqnarray}
 b_{0} (x) =
 -i \;  [\;   \tau_{1}  \; e_{(2)}^{k} \; \partial_{0} e_{(3)k} +
\tau_{2} \; e_{(3)}^{k} \; \partial_{0} e_{(1)k} + \tau_{3} \;
e_{(1)}^{k} \; \partial_{0} e_{(2)k} \;]   \;,
\nonumber
\\
b_{l} (x) = -i\; [\;  \tau_{1}  \; e_{(2)}^{k} \; \nabla_{l}
e_{(3)k} + \tau_{2}\;  e_{(3)}^{k} \; \nabla_{l} e_{(1)k} +
\tau_{3} \; e_{(1)}^{k} \; \nabla_{l} e_{(2)k}\;  ]  \;.
\label{4.6}
\end{eqnarray}

\noindent Therefore eq.(\ref{4.1})  can be rewritten as a system for
constituents  $\Psi (x) = (\Phi_{0}(x), \Phi (x) $, $ E (x) , H(x)
) $:
\begin{eqnarray}
i w^{l}(x) \; ( \nabla_{l} + b_{l}  - ieA_{l}  ) \;  E = m  \;
\Phi_{0} \; ,
\nonumber
\\
-  (  \nabla_{0} + b_{0} - ieA_{0}  ) \; E + i \tau^{l}(x) \; (
\nabla_{l} + b_{l} - ieA_{l}  ) \; H = m \; \Phi \; ,
\nonumber
\\
i \tilde{w}^{l}(x)\;   (\nabla_{l}  - ie A_{l}) \; \Phi_{0} + (
\nabla_{0} + b_{0} - ieA_{0}  ) \;  \Phi = m  \; E \; ,
\nonumber
\\
-i \tau^{l}(x)  \; ( \nabla_{l} + b_{l}  - ieA_{l}  ) \;  \Phi = m
\; H \; , \label{4.7}
\end{eqnarray}

\noindent where
\begin{eqnarray}
\tau^{l}(x) = e^{l}_{(k)}(x)\; \tau^{k} , \qquad w^{l}(x) =
e^{l}_{(k)}(x) \;w^{k}, \qquad \tilde{w}^{l}(x) = e^{l}_{(k)}(x)
\;\tilde{w}^{k} \; .
\nonumber
\end{eqnarray}

\noindent After excluding non-dynamical variables $\Phi_{0}(x)$ and
$H(x)$
\begin{eqnarray}
- ( \nabla_{0} + b_{0} - ieA_{0} ) \; E  +
 i \tau^{l}(x) \; (  \nabla_{l} + b_{l}  - ieA_{l}  ) \; ( -{i
\over m } )\; \tau^{k}(x) \; ( \nabla_{k} + b_{k} - ieA_{k} )  \;
\Phi = m \; \Phi  \; ,
\nonumber
\\
(\nabla_{0} + b_{0} - ieA_{0}  ) \;  \Phi +
 i \tilde{w}^{l}(x) \;  (\nabla_{l}  - ie A_{l}) \;  {i \over m }
\;
 w^{k}(x) \; (  \nabla_{k} + b_{k}  - ieA_{k}  )   \; E
 = m \; E \; .
\nonumber
\\
\label{4.9}
\end{eqnarray}

\noindent Allowing for commutative relations
\begin{eqnarray}
\tau^{k}(x) \; (  \nabla_{l} + b_{l} - ieA_{l}  ) = (  \nabla_{l}
+ b_{l} - ieA_{l} ) \; \tau^{k}(x)\;   ,
\nonumber
\\
(\nabla_{l}  - ie A_{l}) \;  w^{k}(x) = w^{k}(x) \;  (\nabla_{l}
+b_{l}  - ie A_{l}) \; ,
\nonumber
\end{eqnarray}

\noindent eqs. (\ref{4.9}) reduce to ($\bullet$ represent diad
multiplication of vectors)
\begin{eqnarray}
- ( \nabla_{0} + b_{0} - ieA_{0}  ) \; E  +
  {1 \over m } \; \tau^{l}(x) \tau^{k}(x)\; (  \nabla_{l} + b_{l}
- ieA_{l}  )\; (\nabla_{k} + b_{k} - ieA_{k}  ) \;  \Phi = m \;
\Phi  \; ,
\nonumber
\\
+ ( \nabla_{0} + b_{0} - ieA_{0}  )\;  \Phi  -
   {1 \over m } \;  \tilde{w}^{l}(x) \bullet  w^{k}(x) \;
(\nabla_{l}  + b_{l} - ie A_{l})\;
 (  \nabla_{k} + b_{k}  - ieA_{k}  )  \; E    = m \; E \; ,
\nonumber
\end{eqnarray}

\noindent or
\begin{eqnarray}
 D_{0}  E  +
 {1 \over m } \; \tau^{l}(x) \tau^{k}(x) \;   D_{l} D_{k} \;  \Phi = m \;\Phi \; ,
\nonumber
\\
  D_{0}   \Phi -   {1 \over m } \; \tilde{w}^{l} (x) \bullet  w^{k}(x)\;  D_{l} D_{k} \; E
 = m \; E \;.
\label{4.11}
\end{eqnarray}

\noindent Instead of  $\Phi (x)$ and $  E (x) $  let us introduce
$\psi  (x)$ and $ \varphi(x) $:
\begin{eqnarray}
 {1 \over 2} \; ( \Phi + i E ) =\psi  \; , \qquad
{1 \over 2} \; ( \Phi - i E ) = \varphi  \; . \label{4.12}
\end{eqnarray}

\noindent Eqs.  (\ref{4.11})  will look
\begin{eqnarray}
2 m\;  \psi  =  + 2 i D_{0} \; \psi   +
 {1 \over m} \; \tau^{l}(x) \tau^{k}(x)  \;  D_{l} D_{k} \; (\psi
+ \varphi) - {1 \over m}  \; \tilde{w}^{l} (x) \bullet w^{k}(x) \;
D_{l} D_{k} \;  (\psi - \varphi ) \;   ,
\nonumber
\\
2 m \;  \varphi   =  -2 i D_{0} \; \varphi   +
 {1 \over m} \; \tau^{l}(x)  \tau^{k}(x)  \; D_{l} D_{k} \; (\psi
+ \varphi ) + {1 \over m}  \; \tilde{w}^{l}(x) \bullet w^{k}(x) \;
D_{l} D_{k} \;  (\psi - \varphi  ) \;  .
\nonumber
\\
\label{4.13}
\end{eqnarray}

\noindent Making the formal change
 $i D_{0}
\Longrightarrow (iD_{0} + m )$,  we get to
\begin{eqnarray}
0  =  + 2i D_{0}   \;\psi  +
 {1 \over m}\;
\tau^{l}(x) \tau^{k}(x) \; D_{l} D_{k} \; (\psi + \varphi)   -
{1 \over m}   \; \tilde{w}^{l}(x) \bullet  w^{k}(x) \;  D_{l}
D_{k}  \; (\psi - \varphi )  \;  ,
\nonumber
\\
4 m  \; \varphi   =  -2i D_{0} \;  \varphi   +
 {1 \over m}\;
\tau^{l}(x) \tau^{k}(x) \;    D_{l} D_{k} \; (\psi + \varphi)  +
 {1 \over m} \;  \tilde{w}^{l}(x) \bullet  w^{k}(x)\;   D_{l}
D_{k} \; (\psi - \varphi )  \; .
\nonumber
\\
\label{4.14}
\end{eqnarray}

\noindent From eq. (\ref{4.14}), taking $\psi(x)$  as a big component
and $\varphi(x)$  as small we arrive at

\begin{eqnarray}
i D_{0}   \; \psi (x) = {1 \over 2 m}\;  [\; \tilde{w}^{l}(x)
\bullet  w^{k}(x) -  \tau^{l}(x) \tau^{k}(x)\;  ]  \; D_{l} D_{k}
\; \psi (x) \;,
\nonumber
\\
4 m^{2}  \; \varphi (x) = +  \; [\; \tau^{l}(x) \tau^{k}(x)    +
 \tilde{w}^{l}(x) \bullet  w^{k}(x)\; ] \;  D_{l} D_{k} \; \psi (x)  \; .
\label{4.15}
\end{eqnarray}

\noindent From (\ref{4.15}), allowing for identity
$
\tau^{l}(x) \tau^{k}(x)  = - g^{lk}(x) + \tilde{w}^{k}(x) \bullet
w^{l}(x) \;
$ and using notation
\begin{eqnarray}
\tilde{w}^{l}(x) \bullet  w^{k}(x) +  \tilde{w}^{k}(x) \bullet
w^{l}(x) = w^{(lk)}(x) \; ,
\nonumber
\\
 \tilde{w}^{l}(x) \bullet  w^{k}(x) -  \tilde{w}^{k}(x) \bullet  w^{l}(x) = j^{lk}(x)  \;,
\nonumber
\\
 j^{ps} = - i \epsilon _{psj} \tau_{j} , \qquad
j^{lk}(x) = e^{l}_{(p)} e^{k}_{(s)} j^{ps} \; ,
\nonumber
\end{eqnarray}

\noindent  we will obtain
\begin{eqnarray}
4 m^{2}  \; \varphi (x) = ( \; -  D^{l} D_{l}   +   w^{(lk)}(x) \;
D_{l} D_{k} \;) \;  \psi \;  ,
\nonumber
\\
i D_{0}\;  \psi (x) = {1 \over 2 m} \;  (\;  +   D^{l} D_{l}
+   {1\over 2} \; j^{lk}(x) \; [ D_{l}, D_{k} ]_{-}\;   ) \;
\psi (x) \;  .
 \label{4.16}
\end{eqnarray}

Thus, the Pauli equation for 3-vector wave  function in Riemannian
space is (compare with  (\ref{3.18}))
\begin{eqnarray}
i  D_{t}  \; \psi = {1  \over 2 m} \;  ( \;     D^{l} D_{l}  + {1
\over 2} \; j^{lk}(x)\; [D_{l}, D_{k} ]_{-} \; ) \;    \psi  \; .
\label{4.17}
\end{eqnarray}

\noindent One additional point should be stressed. Take notice
that the  non-relativistic wave function is constructed  in terms
of relativistic ones as follows:
\begin{eqnarray}
\psi (x) = {1 \over 2} \; [ \;  \Phi_{i} (x)  + i\; E_{i}(x) \;
]\;,\qquad \mbox{where} \qquad  E_{i}(x) = \Phi_{0i}(x) \; ;
\label{4.18}
\end{eqnarray}

\noindent this function $\psi$ is complex even if  we start with
real-valued relativistic components.

Let us add some details about the term   $ {1 \over 2} j^{lk}(x)\;
[D_{l}, D_{k}]_{-} $ entered (\ref{4.17}):
\begin{eqnarray}
{1 \over 2 } \; j^{lk}(x) \; [ D_{l} , D_{k}]_{-} \; \psi =
  {1 \over 2}  \; j^{lk}(x)  \; ( \;  - i e  F_{lk} +
  \nabla_{l} b_{k} - \nabla_{k} b_{l} + b_{l} b_{k} - b_{k} b_{l}  \; ) \;  \psi  \; .
\nonumber
\end{eqnarray}

\noindent Taking into account relations
\begin{eqnarray}
\nabla_{l} b_{k} - \nabla_{k} b_{l} = + \; j^{dc} \; ( \nabla_{l}
e_{(d)m} ) \;  (\nabla_{k} e ^{m}_{(c)} ) \; +
 {1 \over 2} \; j^{dc} \; e_{(d) m} \; \{ \; \nabla_{l}
\nabla_{k} - \nabla_{k} \nabla_{l}\;  \} \; e_{(c)m}
\nonumber
\end{eqnarray}

\noindent and
\begin{eqnarray}
b_{l} b_{k} - b_{k} b_{l} = {1 \over 4} \;  (  j^{ps}  j^{cd} -
j^{cd}  j^{ps} ) \;  e^{n}_{(p)} \; (\nabla_{l} e_{(s)n} )  \;
e_{(c)}^{m} \; ( \nabla_{k} e_{(d)m} )  =
  -\; j^{dc} \; ( \nabla_{l} e_{(d)m} ) \; (\nabla_{k} e
^{m}_{(c)})\;  ,
\nonumber
\end{eqnarray}

\noindent we find
\begin{eqnarray}
b_{l} b_{k} - b_{k} b_{l} + \nabla_{l} b_{k} - \nabla_{k} b_{l} =
 {1 \over 2}\;  j^{dc} \; e_{(d)}^{m}  \; R_{klmn} \; e_{(c)}^{n}
= {1 \over 2} \; j^{mn}(x) \; R_{lkmn} \; ; \label{4.19}
\end{eqnarray}

\noindent $R_{lkmn}$  is a Riemann curvature tensor for 3-space,
and eq. (\ref{4.17})  can be written as
\begin{eqnarray}
i  D_{t}\;   \psi = {1  \over 2 m} \;  [ \;   D^{l} D_{l} -
 i e \; {1 \over 2}\;
 j^{lk}(x)\;  F_{lk} + {1 \over 4} \;  j^{lk}(x)\;
 j^{mn}(x)\;  R_{lkmn} \; ]
\; \psi  \;  . \label{4.20}
\end{eqnarray}

\noindent In turn, one can readily verify
\begin{eqnarray}
 {1 \over 4}  j^{lk}(x)
\;  j^{mn}(x) R_{lkmn}  =
 {1 \over 4} \;
(-i \epsilon _{prc} \tau_{c} ) \; e^{l}_{(p)}  e^{k}_{(r)}  \;\;
(-i \epsilon_{std} \tau_{d} )\;  e^{m}_{(s)} \; e^{n}_{(t)} \;
R_{lkmn} =
\nonumber
\\
=  -{1 \over 4} \; \vec{\tau}\;  (\; \vec{e}^{\;l} \times
\vec{e}^{\;k} \; ) \;   R_{lkmn} \;  \vec{\tau} \; (\;
\vec{e}^{\;m} \times \vec{e}^{\;n}\; ) \;  . \qquad \qquad  \label{4.21}
\end{eqnarray}

\noindent Thus, the Pauli equation for meson in a curved space
looks as follows (in ordinary units)
\begin{eqnarray}
i \hbar \; D_{t}\;   \psi =- { \hbar^{2}  \over 2 m} \;   [
\;  - D^{l} D_{l} +
 i \; {e \over \hbar c}  \; {1 \over 2}\;
 j^{lk}(x)\;  F_{lk} +
 \nonumber
 \\
   +
  {1 \over 4} \;
\vec{\tau}\;  (\; \vec{e}^{\;l} \times \vec{e}^{\;k} \; ) \;
R_{lkmn} \;  \vec{\tau} \; (\; \vec{e}^{\;m} \times
\vec{e}^{\;n}\;  )
 \;   ]
\; \psi  \;  . \label{4.22b}
\end{eqnarray}

\section{ The wave equation for a scalar particle in Riemannian
space:  non-relativistic approximation }

Now let us turn the case of a scalar particle.
The Klein-Fock-Gordon equation  in a curved space is
\begin{eqnarray}
 [\; (i\hbar \; \nabla_{\alpha} +{e\over c} A_{\alpha}) \;
g^{\alpha \beta}(x) \; (i\hbar \; \nabla_{\beta} +{e\over c}
A_{\beta})
 \; - {\hbar^{2}  \over 6} \; R \;- m^{2} c^{2}  \; ] \; \Psi (x) = 0 \; .
\label{2.9.1a}
\end{eqnarray}

\noindent Take notice on additional interaction term through
scalar curvature $R(x)$  (F. G\"{u}rsey [90]).
\noindent This equation may be changed to the  form more
convenient in application.  With the use of the  known relations [91]
\begin{eqnarray}
\nabla_{\alpha} g^{\alpha \beta} (x) \nabla_{\beta} \; \Phi = {1
\over \sqrt{-g}} {\partial \over \partial x^{\alpha}} \sqrt{-g}
g^{\alpha \beta} {\partial \over \partial x^{\beta} }  \; \Psi \;
, \qquad \nonumber
\\
\nabla_{\alpha} g^{\alpha \beta} A_{\beta} = {1 \over \sqrt{-g}}
{\partial \over \partial x^{\alpha}} \sqrt{-g} g^{\alpha
\beta}A_{\beta} \; , \qquad g= \mbox{det}\; (g_{\alpha \beta} )
\nonumber
\end{eqnarray}

\noindent eq. (\ref{2.9.1a}) is changed to
\begin{eqnarray}
\left  [\; {1 \over \sqrt{-g} }  ( i \hbar {\partial \over
\partial x^{\alpha}} + {e \over c} A_{\alpha} )\; \sqrt{-g}
g^{\alpha \beta } (x) (i\hbar  { \partial \over \partial
x^{\beta}}   + {e \over c} A_{\beta} ) - {\hbar^{2} \over 6} \; R
\;- m^{2}c^{2}   \right  ] \; \Psi (x) = 0 \; . \label{2.9.3c}
\end{eqnarray}

What is the Schr\"{o}dinger's non-relativistic equation in the
curved space-time?

Let us begin with a generally covariant first order equations for a scalar particle
(take notice to the additional interaction term through the Ricci scalar
 \begin{eqnarray}
(i\;  \nabla _{\alpha } +  {e\over c \hbar} A_{\alpha})\; \Phi  =
{mc\over \hbar }\; \Phi _{\alpha }  \; , \qquad
(i\; \nabla _{\alpha } +  {e\over c \hbar } A_{\alpha})\; \Phi
^{\alpha }
 = {mc \over  \hbar } (1 + \sigma  \;{ R(x) \over m^{2}c^{2} / \hbar^{2} } ) \; \Phi \; ,
\label{2.10.1}
\end{eqnarray}

\noindent With the  notation
\begin{eqnarray}
1   + \sigma  \;{ R(x) \over m^{2}c^{2} / \hbar^{2} } = \Gamma (x)
\; . \nonumber
\end{eqnarray}

\noindent eqs.  (\ref{2.10.1}) read
\begin{eqnarray}
(i\;  \partial  _{\alpha } +  {e\over c \hbar} A_{\alpha})\; \Phi
(x)  = {mc\over \hbar }\; \Phi _{\alpha }  \; , \qquad
({i \over \sqrt{-g}} \; {\partial \over  \partial x^{\alpha} }
\sqrt{-g}  + {e \over c \hbar } \; A_{\alpha})\;  g^{\alpha \beta}
\Phi_{\beta} = {mc \over  \hbar }  \; \Gamma \; \Phi \; .
\label{2.10.2}
\end{eqnarray}

In the space-time models  of the  type (\ref{3.1}), one can easily  separate time- and space- variables
in eq. (\ref{2.10.2}):
\begin{eqnarray}
(i\;  \partial  _{0 } +  {e\over c \hbar} A_{0})\; \Phi   =
{mc\over \hbar }\; \Phi _{0 }  \; , \qquad
(i\;  \partial  _{l } +  {e\over c \hbar} A_{l})\; \Phi   =
{mc\over \hbar }\; \Phi _{l}  \; ,  \qquad \nonumber
\\
( i {\partial \over \partial x^{0} }  +  {i \over \sqrt{- g}} \;
{\partial \sqrt{-g} \over  \partial x^{0} } + {e \over c \hbar }
\; A_{0})\;   \Phi_{0} +  ({i \over \sqrt{-g}} \; {\partial \over  \partial x^{k} }
\sqrt{-g}  + {e \over c \hbar } \; A_{k})\;  g^{kl} \Phi_{l}= {mc
\over  \hbar } \; \Gamma \;  \Phi \; . \label{2.10.3}
\end{eqnarray}

\noindent Now one should separate the rest energy - term by means of  the
substitutions:
\begin{eqnarray}
\Phi \Longrightarrow  \exp \; [- i {mc^{2} t \over \hbar } ]\;\;
\Phi \; , \qquad \Phi_{0}   \Longrightarrow  \exp \; [ -i {mc^{2} t
\over \hbar } ]\;\; \Phi_{0} \; , \qquad \Phi_{l} \Longrightarrow
\exp \; [ -i {mc^{2} t \over \hbar } ]\;\; \Phi _{l}\; . \nonumber
\end{eqnarray}

\noindent
As a result, eq.  (\ref{2.10.3})  will give
\begin{eqnarray}
( i \hbar  \;  \partial  _{t }  + mc^{2}  \; + e A_{0})\; \Phi (x)
= mc^{2}\; \Phi _{0 }(x)  \; , \label{2.10.4a}
\\
(  i  \hbar  \;  \partial  _{t }  + mc^{2}    +  i \hbar {1 \over
\sqrt{-g}} \;
 {\partial \sqrt{-g} \over  \partial t }
  + e  \; A_{0})\;   \Phi_{0} +
\nonumber
\\
+ c ({i \hbar  \over \sqrt{-g} } \; {\partial \over  \partial
x^{k} } \sqrt{-g}  + {e \over c}  \; A_{k} )\;  g^{kl} \Phi_{l}=
mc ^{2} \; \Gamma \; \Phi (x)\; , \label{2.10.4b}
\\
(i\; \hbar \; \partial  _{l } +  {e\over c } A_{l})\; \Phi (x)  =
mc \; \Phi _{l}(x)  \; . \label{2.10.4c}
\end{eqnarray}

\noindent With the help of  (\ref{2.10.4c}), the  non-dynamical variable   $\Phi_{l}$
can be readily excluded:
\begin{eqnarray}
( i \hbar  \;  \partial  _{t }  + mc^{2}  \; + e A_{0})\; \Phi (x)
= mc^{2}\; \Phi _{0 }(x)  \; , \qquad \qquad \label{2.10.5a}
\\[2mm]
 ( \;  i  \hbar  \;  \partial  _{t }  + mc^{2}    +  i \hbar
{1 \over \sqrt{-g}} \;
 {\partial \sqrt{-g} \over  \partial t }
  + e  \; A_{0} \;  )\;   \Phi_{0} + \qquad \qquad
\nonumber
\\
+ {1 \over  m} \;  [ \;
 ({i \hbar  \over \sqrt{-g} } \;  \partial_{k}
\sqrt{-g}  + {e \over c}  \; A_{k} )\;
 g^{kl} (i\; \hbar \; \partial  _{l } +  {e\over c } A_{l}) \;  ] \; \Phi (x)=
mc ^{2} \; \Gamma \; \Phi (x)\; . \label{2.10.5b}
\end{eqnarray}

\noindent
Now we are to introduce a small  $\varphi$ and big  $\Psi$ components:
\begin{eqnarray}
\Phi -\Phi_{0} = \varphi , \qquad \Phi + \Phi_{0} = \Psi \; ;
\nonumber
\\
\Phi = {\Psi + \varphi   \over 2} , \qquad \Phi_{0} = {\Psi -
\varphi  \over 2} \;. \label{2.10.6b}
\end{eqnarray}

\noindent Substituting eq.  (\ref{2.10.6b}) into (\ref{2.10.5a})
and  (\ref{2.10.5b}) one  gets
\begin{eqnarray}
( i \hbar  \;  \partial  _{t }  + mc^{2}  \; + e A_{0})\; { \Psi +
\varphi  \over 2}  = mc^{2}\; {\Psi - \varphi   \over 2}   \; ,
\qquad \qquad \label{2.10.7a}
\\
 ( \;  i  \hbar  \;  \partial  _{t }  + mc^{2}    +  i \hbar
{1 \over \sqrt{-g}} \;
 {\partial \sqrt{-g} \over  \partial t }
  + e  \; A_{0} \;  )\;   {\Psi  - \varphi  \over 2}  + \qquad \qquad
\nonumber
\\
+ {1 \over  m} \;  [ \;
 ({i \hbar  \over \sqrt{-g} } \;  \partial_{k}
\sqrt{-g}  + {e \over c}  \; A_{k} )\;
 g^{kl} (i\; \hbar \; \partial  _{l } +  {e\over c } A_{l}) \;  ]
  \;  { \Psi + \varphi \over 2}=
mc ^{2} \; \Gamma \; { \Psi +\varphi \over 2} \; , \label{2.10.7b}
\nonumber
\\\end{eqnarray}

\noindent or after simple calculation   we arrive at

\begin{eqnarray}
( i \hbar  \;  \partial  _{t }  + e A_{0})\; {+\varphi + \Psi
\over 2}  = - mc^{2}\; \varphi    \; , \label{2.10.8a}
\\
( \;  i  \hbar  \;  \partial  _{t }    +  i \hbar {1 \over
\sqrt{-g}} \;
 {\partial \sqrt{-g} \over  \partial t }
  + e  \; A_{0} \; )\;   {\Psi -\varphi  \over 2}  +
\nonumber
\\
+ {1 \over  m} \;  [ \;
 ({i \hbar  \over \sqrt{-g} } \;  \partial_{k}
\sqrt{-g}  + {e \over c}  \; A_{k} )\;
 g^{kl} (i\; \hbar \; \partial  _{l } +  {e\over c } A_{l}) \;  ]
  \;  {\Psi +\varphi  \over 2}=
\nonumber
\\
  =
mc ^{2} \; (\Gamma + 1) \;{\varphi \over 2}   +
  mc ^{2} \; (\Gamma - 1) \;{\Psi \over 2} \; .
\label{2.10.8b}
\end{eqnarray}

In this point, it is better to consider two different cases.
\underline{The first} possibility  is when one  poses an additional requirement
 $\Gamma = 1$, which  means the  absence of the non-minimal interaction term
through $R$-scalar. Then at $\Gamma = 1$, from the previous
equations -- ignoring small component compared with big one -- it
follows
 \begin{eqnarray}
( i \hbar  \;  \partial  _{t }  + e A_{0})\; { \Psi \over 2}  = -
mc^{2}\; \varphi    \; , \hspace{30mm} \label{2.10.9a}
\\
 ( \;  i  \hbar  \;  \partial  _{t }    +  i \hbar {1 \over
\sqrt{-g}} \;
 {\partial \sqrt{-g} \over  \partial t }
  + e  \; A_{0} \;  )\;   {  \Psi \over 2}  + \hspace{20mm}
\nonumber
\\
+ {1 \over  m} \;  [ \;
 ({i \hbar  \over \sqrt{-g} } \;  \partial_{k}
\sqrt{-g}  + {e \over c}  \; A_{k} )\;
 g^{kl} (i\; \hbar \; \partial  _{l } +  {e\over c } A_{l}) \;  ]
  \;  { \Psi \over 2}=
   mc ^{2} \;  \varphi     \; .
\label{2.10.9b}
\end{eqnarray}

\noindent Excluding  the small constituent we arrive at
\begin{eqnarray}
[ \;  i  \hbar  \; ( \partial  _{t }    + {1 \over 2
\sqrt{-g}} \;
 {\partial \sqrt{-g} \over  \partial t })
  + e  \; A_{0} \;  ] \;   \Psi =
 \nonumber
 \\
  =
  {1 \over 2m}\;
  [ \;
 ({i \hbar  \over \sqrt{-g} } \;  \partial_{k}
\sqrt{-g}  + {e \over c}  \; A_{k} )\; (- g^{kl})\;  (i\; \hbar \;
\partial  _{l } +  {e\over c } A_{l}) \;  ]
  \;  \Psi
\label{2.10.10a}
\end{eqnarray}

\noindent With the  help of substitution
$ \Psi  \;  \Longrightarrow \;
(-g)^{-1/4} \; \Psi $  the obtained equation can be simplified:

\begin{eqnarray}
( \;  i  \hbar  \;  \partial  _{t }   + e  \; A_{0} \; ) \;   \Psi
= \hspace{30mm} \nonumber
\\
= {1 \over 2m}\;  [ \; ({i \hbar  \over \sqrt{-g} } \;
\partial_{k} \sqrt{-g}  + {e \over c}  \; A_{k} )\; (- g^{kl}
)\;(i\; \hbar \; \partial  _{l } +  {e\over c } A_{l}) \;  ]
  \;  \Psi \; ,
\label{2.10.10b}
\end{eqnarray}

\noindent which is the  the Schr\"{o}dinger equation in curved space.

\underline{The second} possibility is  when $\Gamma \neq 1$, then from (\ref{2.10.8b})
\begin{eqnarray}
( i \hbar  \;  \partial  _{t }  + e A_{0})\; { \Psi
\over 2}  = - mc^{2}\; \varphi    \; , \label{2.10.8a'}
\\
( \;  i  \hbar  \;  \partial  _{t }    +  i \hbar {1 \over
\sqrt{-g}} \;
 {\partial \sqrt{-g} \over  \partial t }
  + e  \; A_{0} \; )\;   {\Psi  \over 2}  +
\nonumber
\\
+ {1 \over  m} \;  [ \;
 ({i \hbar  \over \sqrt{-g} } \;  \partial_{k}
\sqrt{-g}  + {e \over c}  \; A_{k} )\;
 g^{kl} (i\; \hbar \; \partial  _{l } +  {e\over c } A_{l}) \;  ]
  \;  {\Psi \over 2}=
\nonumber
\\
  =
mc ^{2} \; (\Gamma + 1) \;{\varphi \over 2}   +
  mc ^{2} \; (\Gamma - 1) \;{\Psi \over 2} \; .
\label{2.10.8b'}
\end{eqnarray}

\noindent
With the use of (\ref{2.10.8a'}) we can derive the following equation for the big
component $\Psi$:
\begin{eqnarray}
( \;  i  \hbar  \;  \partial  _{t }    +  i \hbar {1 \over
\sqrt{-g}} \;
 {\partial \sqrt{-g} \over  \partial t }
  + e  \; A_{0} \; )\;   {\Psi  \over 2}
  +
  \nonumber
  \\
  +
   {(\Gamma + 1)\over 2}  \; ( i \hbar  \;  \partial  _{t }  + e A_{0})\; { \Psi
\over 2} \;     -
  mc ^{2} \; (\Gamma - 1) \;{\Psi \over 2}
  =
\nonumber
\\
- {1 \over 2 m} \;  [ \;
 ({i \hbar  \over \sqrt{-g} } \;  \partial_{k}
\sqrt{-g}  + {e \over c}  \; A_{k} )\;
 g^{kl} (i\; \hbar \; \partial  _{l } +  {e\over c } A_{l}) \;  ]
  \;  \Psi .
\label{2.10.8b''}
\end{eqnarray}

\noindent
This equation can be rewritten as follows:
\begin{eqnarray}
[ \; ({1\over 2} + {1\over 2}\; {\Gamma (x)  +1 \over 2})
( i  \hbar  \partial  _{t }   + e  \; A_{0}) + {i \hbar \over 2
\sqrt{-g}} \;
 {\partial \sqrt{-g} \over  \partial t })
   \;  ] \;   \Psi =
 \nonumber
 \\
  =
  {1 \over 2m}\;
  [ \;
 ({i \hbar  \over \sqrt{-g} } \;  \partial_{k}
\sqrt{-g}  + {e \over c}  \; A_{k} )\; (- g^{kl})\;  (i\; \hbar \;
\partial  _{l } +  {e\over c } A_{l}) \;  ]
  \;  \Psi + mc ^{2} \; {(\Gamma  (x) - 1)\over 2} \; \Psi
\label{2.10.10'}
\end{eqnarray}

\noindent
It remains  to recall that
$$
\Gamma (x)  = 1 + {1 \over 6}\;  {   \hbar^{2} R (x) \over m^{2}c^{2} } \; ,
$$

\noindent
so  the previous equation will take the form
\begin{eqnarray}
\left [ \;  ( 1 +  {1 \over 24}\;   { \hbar^{2} R (x)  \over  m^{2}c^{2} })\;
( i  \hbar  \partial  _{t }   + e  \; A_{0}) + {i \hbar \over 2
\sqrt{-g}} \;
 {\partial \sqrt{-g} \over  \partial t })
   \; \right ] \;      \Psi =
 \nonumber
 \\
  =
  {1 \over 2m}\;
  [ \;
 ({i \hbar  \over \sqrt{-g} } \;  \partial_{k}
\sqrt{-g}  + {e \over c}  \; A_{k} )\; (- g^{kl})\;  (i\; \hbar \;
\partial  _{l } +  {e\over c } A_{l}) \;  ]
  \;  \Psi + mc ^{2} \; {  \hbar^{2} R \over 12 m^{2}c^{2} } \; \Psi \; ,
  \nonumber
\end{eqnarray}

\noindent and finally
\begin{eqnarray}
\left [ \;  ( 1 +   {1 \over 24}\;  { \hbar^{2}  R (x) \over  m^{2}c^{2} })\;
( i  \hbar  \partial  _{t }   + e  \; A_{0}) + {i \hbar \over 2
\sqrt{-g}} \;
 {\partial \sqrt{-g} \over  \partial t } \; \right ]
   \;     \Psi =
 \nonumber
 \\
  =
  {1 \over 2m}\;
  \left [ \;
 ({i \hbar  \over \sqrt{-g} } \;  \partial_{k}
\sqrt{-g}  + {e \over c}  \; A_{k} )\; (- g^{kl})\;  (i\; \hbar \;
\partial  _{l } +  {e\over c } A_{l}) \;  ]
  \;  +   \hbar^{2}  {   R \over  6 } \right ]  \; \Psi \; ,
\end{eqnarray}

\noindent which should be considered as a
 Schr\"{o}dinger equation in  a space-time with non-vanishing scalar curvature
 $R(x) \neq 0$ when allowing for a non-minimal interaction term through scalar curvature $R(x)$.

In addition, several  general comments may be given.
The  wave function of Schr\"{o}dinger equation  $\Psi$ does not coincide
with the initial scalar Klei-Fock  wave function $\Phi$. Instead we have the  following
\begin{eqnarray}
\Psi = \Phi + \Phi_{0} , \qquad \Phi_{0} \;\; \mbox{belongs to} \;\;  \{ \;  \Phi_{0},
\Phi_{1} , \Phi_{2}, \Phi_{3} \; \}\;. \label{2.10.11}
\end{eqnarray}

One may have looked at this fact as  a non-occasional an even necessary  one (in this context see
recent discussion of the problem in [92,93]).
Indeed, let one start with a neutral scalar particle theory. Such a particle
cannot interact with electromagnetic field and its wave function is real. However,
by general consideration, certain non-relativistic limit in this theory must exist.
It  is the case in fact:   the  added term in
(\ref{2.10.11})
\begin{eqnarray}
\Phi_{0} = i {\hbar \over mc }\;  \partial_{0}  \Phi \;
\end{eqnarray}

\noindent is imaginary even if  $ \Phi^{*} = + \Phi $.
All the more, that situation is in accordance with the
 the mathematical structure of the Schr\"{o}dinger equation itself, it cannot be written
 for real wave function at all.

The same property was  seen in the theory of a vector particle:
even if  the  wave function of the relativistic particle of spin 1 is  taken real, the corresponding
wave function in  the non-relativistic approximation turn to be  complex-valued.
By general consideration, one may expect an analogous result in the theory of a spin 1/2
 particle:
if the non relativistic approximation is done in the theory of Majorana neutral
particle  [94] with the real 4-spinor wave function then the corresponding
Pauli spinor wave function must be complex-valued.

This  work was  supported  by Fund for Basic Research of Belarus
and JINR  F06D-006.
 Authors are  grateful  to  participants of
seminar of Laboratory of Physics of Fundamental Interaction,
 National Academy of Sciences of Belarus, for discussion and  advice.

\end{document}